\documentclass[twocolumn,pra,aps,unsortedaddress,showpacs]{revtex4-1}
\usepackage{url,psfrag,graphicx}
\usepackage{placeins}
\usepackage{dcolumn}
\usepackage{amsmath,amssymb}
\usepackage{bm}
\usepackage{pstricks}
\usepackage{hyperref}
\usepackage{epsfig}

\def\<{\langle}
\def\>{\rangle}
\def\ket|#1>{\left|#1\right>}
\def\bra<#1|{\left<#1\right|}
\def\elem<#1|#2|#3>{\left<#1\right|#2\left|#3\right>}
\def\braket<#1|#2>{\langle#1|#2\rangle}
\def\({\left(}
\def\){\right)}
\def\[{\left[}
\def\]{\right]}
\def\{{\lbrace}
\def\}{\rbrace}
\def\beq{\begin{equation}}
\def\eeq{\end{equation}}
\def\barray{\begin{eqnarray}}
\def\earray{\end{eqnarray}}

\def\C{{\mathbb C}}
\def\myi{\imath}
\def\sx{\sigma^{x}}
\def\sy{\sigma^{y}}
\def\sz{\sigma^{z}}
\def\gs{ \ket|{\rm g}>}
\def\brags{\<{\rm g} |}

\begin{document}

\title[Short Title]{Fourier-space entanglement of spin chains}

\author{Miguel Ib\'a\~nez-Berganza}
\affiliation{INFN-Gruppo Collegato di Parma, via G.P. Usberti, 7/A,
  43124, Parma,  Italy \\
Dipartimento di Fisica, Universit\`a di Roma ``La Sapienza'', Piazzale
Aldo Moro 5, 00185, Roma,  Italy}
\author{Javier Rodr\'{\i}guez-Laguna}
\affiliation{Dept. of Fundamental Physics, Universidad Nacional de
  Educaci\'on a Distancia (UNED), Madrid, Spain}
\author{Germ\'an Sierra}
\affiliation{Instituto de F\'{i}sica Te\'orica (IFT), UAM-CSIC, Madrid, Spain}

\date{February 12, 2016}

\begin{abstract}
Entanglement between different regions in momentum space is studied
for ground states of some spin-chain Hamiltonians: the XY model, the
Ising model in a transverse field (ITF) and the XXZ models. In the XY
and ITF cases, entanglement only takes place between states with
opposite momenta. Thus, an anisotropy in the interaction induces
entanglement in the momentum pairs. In the ITF case, the ferromagnetic
phase is characterized by a total entropy between left- and
right-moving modes which is independent on the external field. This
result characterizes the Ising phase transition in momentum space. In
the critical XXZ case, we provide evidence that the maximal entropy
between energy modes around the Fermi point grows logarithmically with
the system size, with a prefactor which depends on the
compactification radius. The slow growth of the entanglement in
Fourier space with the system size provides an explanation for the
success of the renormalization techniques in momentum space.
\end{abstract}

\maketitle


\section{Introduction}

Entanglement entropy (EE) and entanglement spectra (ES) are concepts
in quantum information theory which have provided invaluable insight
for quantum phase transitions and topological phases
\cite{Amico.RMP.08,Li_Haldane.PRL.08,Wen.PRB.10}. Area laws
\cite{Eisert.RMP.10,Wolf.08} have been put forward for the ground
state (GS) of many local Hamiltonians, stating that the EE of a block
in real space is proportional to the number of links which must be
broken in order to isolate the subsystem. A rigorous proof can be given
in few situations, such as 1D gapped systems
\cite{Hastings.JSTAT.07}. The EE for 1D critical states is known to
receive a logarithmic correction proportional to the central charge of
the associated conformal field theory (CFT)
\cite{Vidal_Latorre.PRL.03,Calabrese.JSTAT.04}. The fact that
entanglement is a resource for quantum computation
\cite{Nielsen_Chuang} can be reversed, to state that it is the
limiting feature for many numerical methods designed to study quantum
many-body problems \cite{Vidal.PRL.03}, such as the density matrix
renormalizartion group (DMRG) \cite{White.93}.

Most studies of entanglement are performed in real space. Yet, many
analytical and numerical methods for quantum many-body systems rely on
a Fourier space description. Real space sites and momentum sites
constitute two different tensor structures of the same many-body
Hilbert space, thus allowing for very different entanglement
patterns. The success of DMRG in momentum space \cite{Xiang.PRB.96}
and the relation between momentum-space entanglement and
renormalizability \cite{Balasubramanian.11} suggest that the
dependence of the maximum entanglement with the system size could be
lower in momentum space than in real space for many systems of
interest. A duality between entanglement in real and momentum space
was uncovered in \cite{Klich.PRL.06,Lee.14}, which applies to free
fermionic systems, but no general rule has been derived. Thus, while
\cite{Lauchli.14} reports that the XXZ critical region does not have
any signatures in the ES or the EE, \cite{Laguna.PRB.14} shows that
the ES in energy space can provide a unified picture of the phase
diagram for $p$-wave superconductors, and \cite{Anfossi.PRB.08} finds
similar traces of the transition of the extended Hubbard model
studying multipartite entanglement in momentum space.

For a homogeneous system, the entanglement properties of a spatial
block should only depend on the block size. Fourier space, on the
other hand, is inherently inhomogeneous. For systems with a Fermi
surface structure, the Fermi momentum $k_F$ is specially
relevant. Thus, it makes special sense to study the entanglement
between blocks around $k_F$ and their complementary, and the
entanglement between levels below and above $k_F$ \cite{Lee.14}, which
has proved valuable in the study of superconducting systems
\cite{Dukelsky.99,Laguna.PRB.14}. As we will see, it is thus crucial
to know the nature of our physical problem before deciding which
blocks might provide a deeper insight.

This article is organized as follows. A general discussion about the
meaning of entanglement in Fourier space is done in section
\ref{sec:mbft}. Then, section \ref{sec:xy} shows the behaviour of the
entanglement in momentum space for the (generalized) XY model,
computed analytically via a Jordan-Wigner transformation. Section
\ref{sec:xxz} is devoted to the (numerical) study of the entanglement
in Fourier space of the XXZ model. The article ends with a section on
conclusions and proposals for further work.


\section{Entanglement from real to momentum space}
\label{sec:mbft}

Let us consider a system of spinless fermions. Its Hilbert space is
the tensor product of $N$ qubit Hilbert spaces: ${\cal H}=\C^2 \otimes
\cdots \otimes \C^2$. A basis can be written using the local creation
operators $c^\dagger_i$:

\begin{equation}
\ket|n_1\cdots n_N> = \prod_{i=1}^N (c^\dagger_i)^{n_i} \ket|0>,
\label{eq:real_space_state}
\end{equation}
where $n_i\in\{0,1\}$. For any single-body unitary transformation $U$,
we can define a new set of creation operators: $b^\dagger_j = \sum_i
U_{ji} c^\dagger_i$. Now we can define a new basis:

\begin{equation}
\ket|m_1\cdots m_N> = \prod_{j=1}^N (b^\dagger_j)^{m_j} \ket|0>,
\label{eq:momentum_space_state}
\end{equation}
with $m_i\in \{0,1\}$. Defining $\vec n\equiv\{n_1,\cdots,n_N\}$ and
$\vec m\equiv\{m_1,\cdots,m_N\}$, any pure state of ${\cal H}$ can be
expressed in both bases:

\begin{equation}
\ket|\Psi>=\sum_{\vec n} C_{\vec n} \ket|n_1\cdots n_N> =
\sum_{\vec m} B_{\vec m} \ket|m_1\cdots m_N>.
\label{eq:pure_state_both_basis}
\end{equation}
And we can write a change of basis matrix:

\begin{equation}
B_{\vec m}= \sum_{\vec n} \Omega_{\vec m,\vec n}\; C_{\vec n},
\label{eq:def_basis_change}
\end{equation}
where $\Omega_{\vec m,\vec n}=\bra<m_1\cdots m_N|n_1\cdots n_N\>$. In
the particular case of fermionic states, both basis states are Slater
determinants, so the scalar products can be computed efficiently:

\begin{equation}
\Omega_{\vec m, \vec n}=\det\( U_{\vec m, \vec n} \),
\label{eq:basis_change}
\end{equation}
where we use the notation $U_{\vec m, \vec n}$ to be the (sub-)matrix
where we only pick the rows given by the values $m_j=1$ and the
columns with $n_i=1$.

When the single-body unitary matrix $U$ implements the Fourier
transform, then the coefficients $B_{\vec m}$ are read as the Fourier
expansion of the many-body wavefunction, which correspond to a
different {\em slicing} of the Hilbert space as a tensor product: each
qubit space corresponds now to a {\em momentum site} instead of a real
space site. Each index $j$ corresponds to a certain momentum $k_j$,
determined by the size of the system and the boundary conditions. We
will focus on antiperiodic boundary conditions with even $N$. Thus, we
have $k=(2p+1)\pi/N$, with integer $p$. This way, $k=0$ and $k=\pi$
are never allowed. We always define our momenta to lie in
$(-\pi,\pi]$.

If a quantum single-body state is translationally invariant, it must
have a well defined momentum. A translationally invariant quantum
many-body state fulfills a more involved constraint: its non-zero
amplitudes $B_{\vec m}$ all have $\sum_j m_j k_j=0$ mod $2\pi$.

\begin{figure}[h!]
\epsfig{file=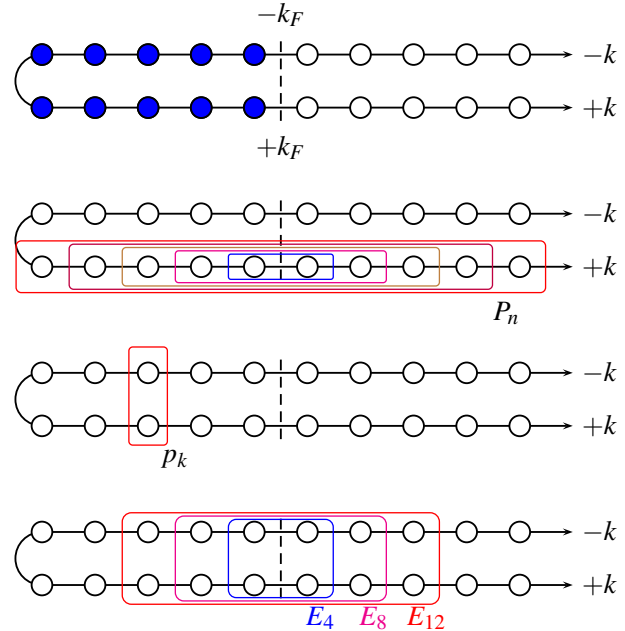,width=8cm}
\caption{\label{fig:illust}Illustration of the different blocks used
  to study the entanglement entropy in momentum space. (A) momentum
  space structure, showing in blue the momenta occupied by a Fermi
  sea. (B) Positive momenta blocks, $P_n$ contains the $n$ momenta
  closest to the positive Fermi level, $+k_F$. (C) Pairing of opposite
  momenta, block $p_k$ contains the momenta $k$ and $-k$. (D) Energy
  blocks, $E_n$ contains the $n$ momenta which are closest to the
  Fermi energy.}
\end{figure}

Momentum space is inherently non-homogeneous, even if real space can
be considered to be so. When we study entanglement in real space,
typically we are only interested in the size of the block. But in
momentum space, we are also interested in the position of those
blocks. Thus, we have defined a few interesting possibilities,
motivated by the physics of the problems that we will study, depicted
in Fig. \ref{fig:illust}.

Let us start our description with a Fermi sea, as depicted in
Fig. \ref{fig:illust} (A), where the Fermi momentum $\pm k_F$ is
marked. The first type of blocks to be considered is $P_n$, which
contains $n$ {\em positive} momenta around the positive Fermi point,
see Fig. \ref{fig:illust} (B). Thus, $P_{N/2}$ corresponds to the
block which contains all positive momenta, which we will denote simply
by $P$. Another natural set of blocks can be formed by choosing the
$(-k,k)$ pairs with the same energy, which we will call $p_k$, as
shown in Fig. \ref{fig:illust} (C). Following \cite{Laguna.PRB.14}, we
can also consider blocks in {\em energy space}, assuming a relation
$E\sim |k|$, as in Fig. \ref{fig:illust} (D), where $E_n$ contains the
$n$ momenta whose energy are closest to the Fermi energy.


\section{Generalized XY model}
\label{sec:xy}

In this section we will present our first physical system, the 1D XY
model, and map it into a model of spinless fermions. The XY
Hamiltonian in a one-dimensional lattice of $N$ sites and periodic
boundary conditions can be written as:

\begin{equation}
H= \frac{-1}{2}\sum_{j=1}^N\frac{1}{2} \left[(1+\gamma)\sx_j\sx_{j+1}
  + (1-\gamma)\sy_j\sy_{j+1}\right] + J\sz_j.
\label{Hxy}
\end{equation}

This system is exactly solvable \cite{Lieb.61}. The procedure is to
perform a Jordan-Wigner transformation, which will bring it to a
fermionic Hamiltonian without particle preservation, followed by a
Fourier transform of the resulting fermionic operators. One obtains
the following Hamiltonian:

\begin{equation}
{\cal H}=\sum_{j\in\Omega}\, A_j\, d_j^\dag\,d_j + i\frac{B_j}{2} \, 
\left[d_j^\dag d_{-j}^\dag+ d_j d_{-j}\right],
\label{XYmomentum1}
\end{equation}
where we defined $A_j$, $B_j$:

\begin{equation}
A_j=J-\cos{k_j}, \qquad B_j=-\gamma \sin{k_j}.
\end{equation}
and where $k_j=2\pi j/N$, and the set of allowed momenta is
$\Omega=\{k_j\}_{j=-N/2}^{N/2}$. The operators $d_j$ are Fourier-space
fermionic operators annihilating a fermion with momentum $k_j$. In the
fermionic formulation, the total $z$-axis magnetization
$M^{(z)}=\sum_j\<\sz_j\>$ of the spin model is related with the number
of fermions, $n_{\rm f}=\sum_j\<d_j^\dag d_j\>$ as: $2n_{\rm
  f}-N=M^{(z)}$.  The following step is a Bogoliubov transformation of
the form

\begin{equation}
b_j^\dag=u_j\, d_j^\dag+ \myi v_j\, d_{-j}, \nonumber
\label{Bogolubovt}
\end{equation}
with $u_j^2+v_j^2=1$ and 

\begin{eqnarray}
u_j^2-v_j^2=A_j/E_j, \nonumber  \\
2u_j v_j = B_j/E_j,
\label{Bogoluboveqns}
\end{eqnarray}
where $E_j$ is the energy of the $j$-th free mode, satisfying
$E_j^2=A_j^2+B_j^2$. Thus,

\begin{equation}
u^2_j= {1\over 2}\( 1+ {A_j\over E_j}\), \qquad
v^2_j= {1\over 2}\( 1- {A_j\over E_j}\).
\label{isolateuv}
\end{equation}

Such a transformation brings the Hamiltonian into a free Hamiltonian
in the $b$-modes:

\begin{equation}
{\cal H} = \sum_j |E_j| b^\dagger_j b_j - \frac{1}{2} \sum_j  E_j.
\label{H_free}
\end{equation}

For each eigenstate of the XY Hamiltonian \eqref{Hxy}, there exists a
corresponding eigenstate of \eqref{H_free} with equal energy, of the
form:

\begin{equation}
\ket|\theta>=
\prod_j \[ \theta_j b_j^\dag + (1-\theta_j) \] \gs,
\label{fermionicstate}
\end{equation}
where $\gs$ is annihilated by all $b_j$'s, and the state is
characterized by the set of occupations $\theta_j=0,1$ in terms of
$b$-modes: $\theta_j=\bra<\theta| b^\dag_j b_j \ket|\theta>$. The
Hamiltonian ground state $\gs$ is related with the $d$'s zero in the
following form:

\begin{equation}
\gs = \prod_j b_j \ket|0>.
\label{zerosrelation}
\end{equation}

Taking $\gamma=0$ (XX model), the Hamiltonian \eqref{XYmomentum1} is
already diagonal in the $d$-modes, so each $d_j$ is correlated only
with itself, $\<d_j^\dag\,d_{j'}\>=\kappa_j \delta_{j,j'}$, where
$\kappa_j=0,1$ is the occupation of site $j$, and where the expected
value $\< \cdot \>$ is taken with respect to the $d$ operator vacuum,
$\ket|0>$. In other words, the eigenstates of the Hamiltonian are
product states in momentum space, hence presenting a vanishing
entanglement entropy in momentum space. For $\gamma\neq 0$, however,
the eigenstates \eqref{fermionicstate} are such that {\em moment $k$
  is only entangled to moment $-k$}. Let us call $(k,-k)$ a {\em
  momentum pair}. Given any block $B$ in momentum space, we need only
to consider its broken pairs: if both $k$ and $-k$ belong to $B$, then
they do not contribute to entanglement. Thus, without any loss of
generality. let us only consider in the following blocks composed of
{\em positive momenta}, see Fig. \ref{fig:illust}.

The von Neumann and R\'enyi entropies quantifying the amount of
entanglement between $B$ and its complementary can be exactly computed
throught the fermionic correlators $\<d^\dag_i d_j\>$ of the
model. Consider the $\ell\times\ell$ {\em reduced correlation matrix},
$\cal C$:

\beq
{\cal C}_{j,j'} = 
\bra<\theta |d_j^\dag d_{j'}\ket|\theta> \qquad 
k_j,k_{j'} \in {\cal A}_\ell.
\eeq
Using the inverse Bogoliubov transformation \eqref{Bogolubovt}, it is
found to be

\barray
{\cal C}_{i,j}=\delta_{i,j} \lambda_j, \nonumber \\
\lambda_j \equiv \[ u_j^2 \theta_j + v_j^2 (1-\theta_{-j}) \].
\label{rcm}
\earray

Let us denote by $\rho_B$ the reduced density matrix of the state
\eqref{fermionicstate} in $B$, i.e., with the degrees of freedom of
the complementary of $B$ traced out. It is a tensor product of
$2\times 2$ reduced density matrices in different sites of $B$:
$\rho_B=\bigotimes_{j\in B} \varrho_j$. The entropy of entanglement is
obtained from the eigenvalues of $\varrho_j$,
$\{\lambda_j,1-\lambda_j\}$, being $\lambda_j$ the eigenvalue of the
$j$-site reduced correlation matrix \eqref{rcm}. The von Neumann
entropy is given by

\beq
S[\rho_B] = \sum_{k_j\in B} H_2(\lambda_j),
\label{Smomentum}
\eeq
where $H_2(x)\equiv -\(x\log x+(1-x)\log (1-x)\)$. Equivalently, the
R\'enyi entropy of order $\alpha>1$ reads:

\beq 
S_\alpha [\rho_B] = {1\over 1-\alpha} \sum_{k_j\in B} \log\[\lambda_j^\alpha +
(1-\lambda_j)^\alpha\].
\label{Snmomentum}
\eeq

Alternatively, one can obtain Eqs. \eqref{Smomentum} and
\eqref{Snmomentum} computing explicitly the reduced $j$-site
correlation matrix using Eq. \eqref{zerosrelation}; in the
$\theta_j=0$ case, for instance, it is simply:
$\varrho_j=v_j^2d_j^\dag \ket|0>_j\bra<0|d_j +
u_j^2\ket|0>_j\bra<0|$. Assuming {\em even} states 
$\theta_j=\theta_{-j}$, the entropies do not depend on the state
$\theta$:

\begin{eqnarray}
S[\rho_B] &=& \sum_{k_j\in B} H_2(u_j^2), \nonumber\\
S_\alpha[\rho_B] &=& {1\over 1-\alpha} \sum_{k_j\in B} 
\log\left[u_j^{2\alpha} + v_j^{2\alpha}\right] \qquad \alpha>1.
\label{Smomentumgeneral}
\end{eqnarray}

Our preferred blocks to show the structure of the entanglement in the
XY model will be the positive momentum blocks centered on the Fermi
point, as illustrated in \ref{fig:illust} (B), i.e., the $P_n$ blocks.

\subsection{XY model with $J=0$}

\begin{figure}[t!]                        
\begin{center} 
\epsfig{file=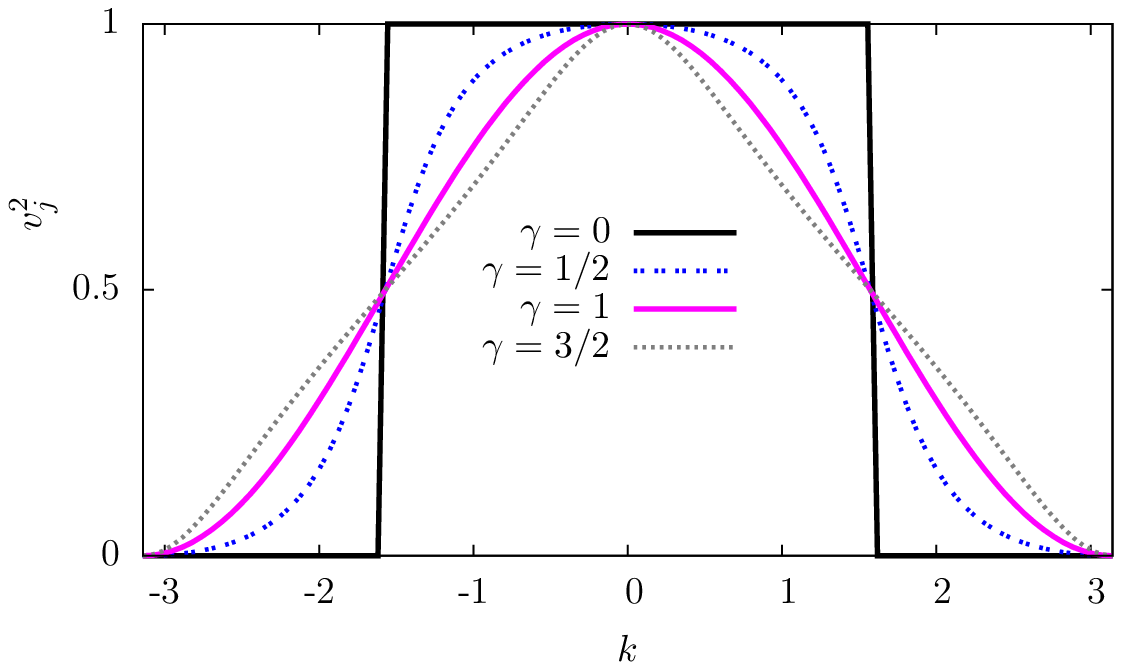,width=0.98\columnwidth}
\epsfig{file=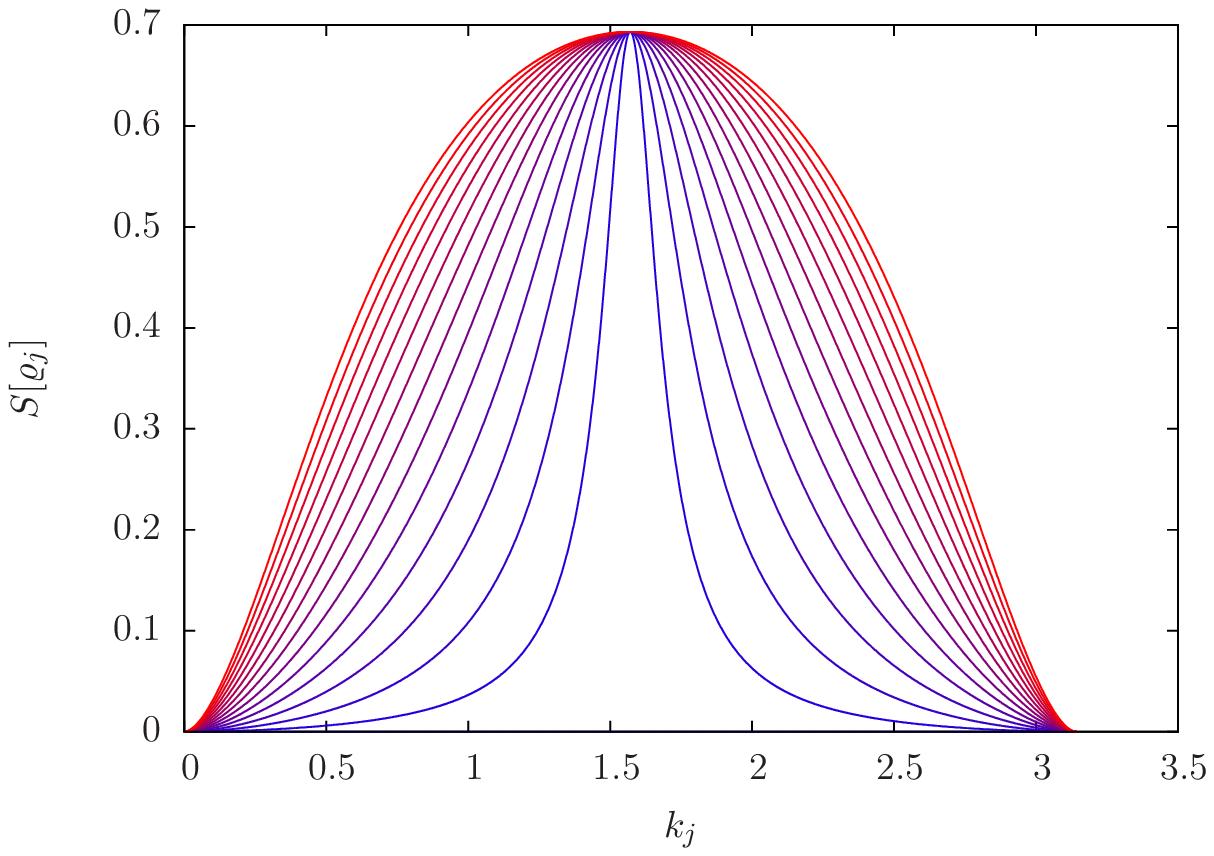,width=8cm}
\epsfig{file=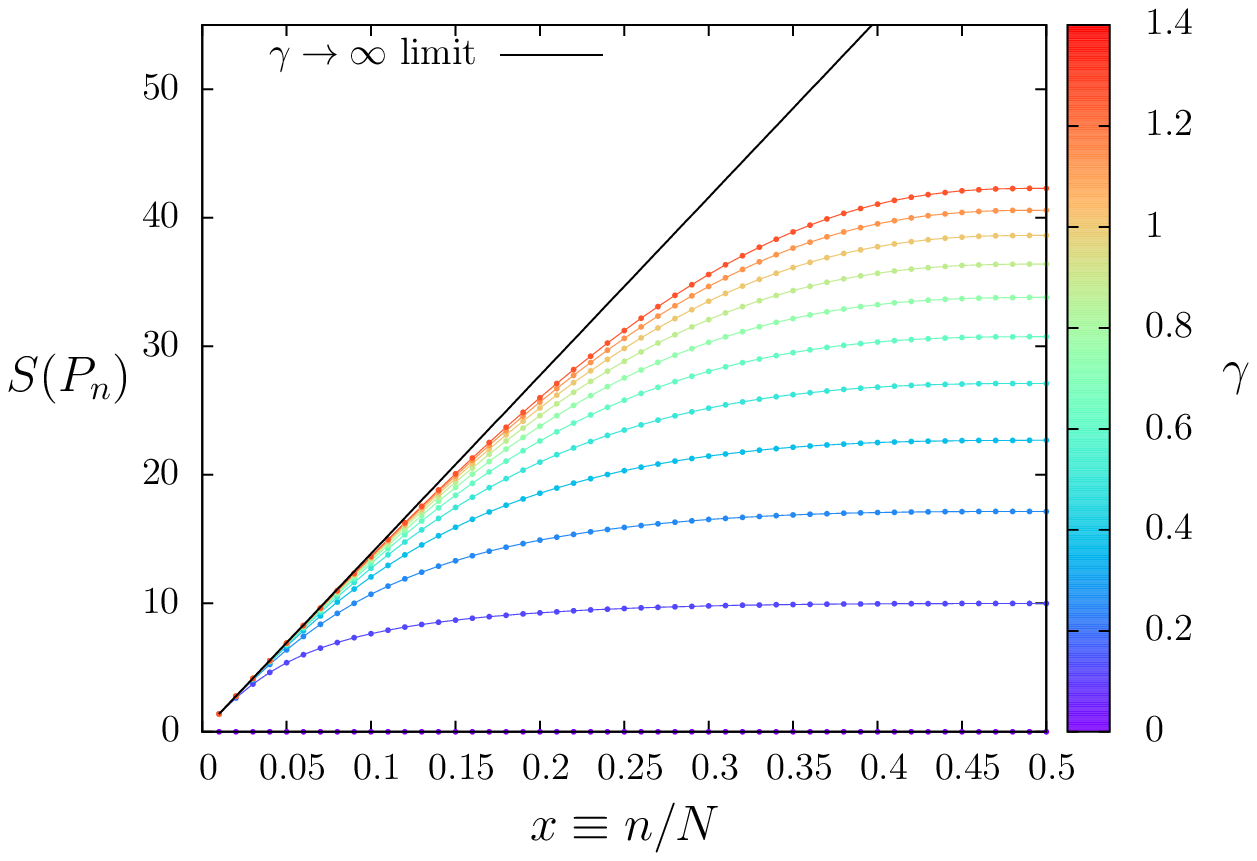,width=0.98\columnwidth}
\caption{\label{fig:gamma}Entanglement in momentum space for the XY
  model of Eq. \eqref{Hxy} with $J=0$, for different values of
  $\gamma$. (A) Occupations of the different modes, $\brags d_j^\dag
  d_j \gs = v_j^2$ vs. $k_j$. (B) Contribution of each mode $k$ to the
  entanglement entropy, $S[\varrho_j]$, with $\gamma$ ranging from 0
  (identically zero) to $1.5$ (red). (C) Entanglement entropy of
  positive momentum blocks $S(P_n)$ as a function of the block size
  $x=n/N$ with $N=100$, along with the $\gamma\to\infty$ limit,
  $S(P_n)=n \log 2$. Again, for $\gamma=0$ the entropy is identically
  zero.}
\end{center}   
\end{figure}

Let us illustrate the laws \eqref{Smomentumgeneral} in their
particular realizations for the XY model ground states. Our first
example is the $J=0$ case. The model with $J=0$ is gapped except for
the point $\gamma=0$ which corresponds to the XX model, whose ground
state is a product state in momentum space. The $\gamma$ parameter may
be regarded as a {\em mass} term. For $\gamma=0$, the state is

\begin{equation}
\gs=\prod_{|k_j|<\pi/2} d^\dag_j \ket|0>.
\end{equation}
The diagonal of matrix $\cal C$ in Eq. \eqref{rcm} gives the
occupations of $d$ modes in the ground state, $\brags d^\dag_j d_j
\gs=v_j^2$, which are presented in Fig. \ref{fig:gamma} (A).

We also present the entanglement entropy of the single momentum
states, $S[\varrho_j]$, which are given by each one of the terms in
the sum of Eqs. (\ref{Smomentum},
\ref{Snmomentum}). Fig. \ref{fig:gamma} (B) shows $S[\varrho_j]$ for
$j=0,\dots,N/2$ as a function of both $k_j$ and $\gamma$. While for
small values of $\gamma$ only those momenta immediately around the
Fermi point at $k_F=\pi/2$ present a significant amount of
entanglement, as $\gamma$ becomes larger and larger the distribution
of entanglement $S[\varrho_j]$ as a function of $k_j-k_F$ becomes
broader.

In Fig. \ref{fig:gamma} (C), we present the $J=0$ entropy $S(P_n)$ as
a function of $n=0,\dots,N/2$ for 11 different values of $\gamma$ from
0 to 1.4 and $N=100$. For $\gamma=0$ one also observes a vanishing
entropy for all $n$, as already mentioned. Let us remark that the
critical case presents the minimum momentum space entropy for all $n$,
as opposed to the behaviour of the real-space entanglement
entropy. For increasing $\gamma$ we get increasing entropy. The
$\gamma\to\infty$ case presents a linear growth of the entropy with
the system size, $S(P_n)\to n \log 2$, as can be analytically proved
from the exact solution.

Notice that, for $\gamma\to\infty$, each pair $(-k,k)$ becomes
maximally entangled, contributing $\log(2)$ to the entropy of the
block of positive momenta. Thus, the entanglement entropy of the block
$P_{N/2}$ tends to its maximal possible value, $N\log(2)/2$. Although
it is tempting to think of this state as a highly entangled state, we
should take into account that it depends on the neighborhood structure
that we impose on momentum space. If we consider $k$ to be neighbor to
$-k$, then the entanglement is only of short range.

\subsection{Ising model}

Let us now study the Ising model, Eq. \eqref{Hxy} with $\gamma=1$, for
different values of $J$. The system is gapped for all $J\neq 1$, and
for $J=1$ it presents a quantum phase transition. The $J<1$ phase is
ferromagnetic, while for $J>1$ the system is paramagnetic.

\begin{figure}[t!]                        
\epsfig{file=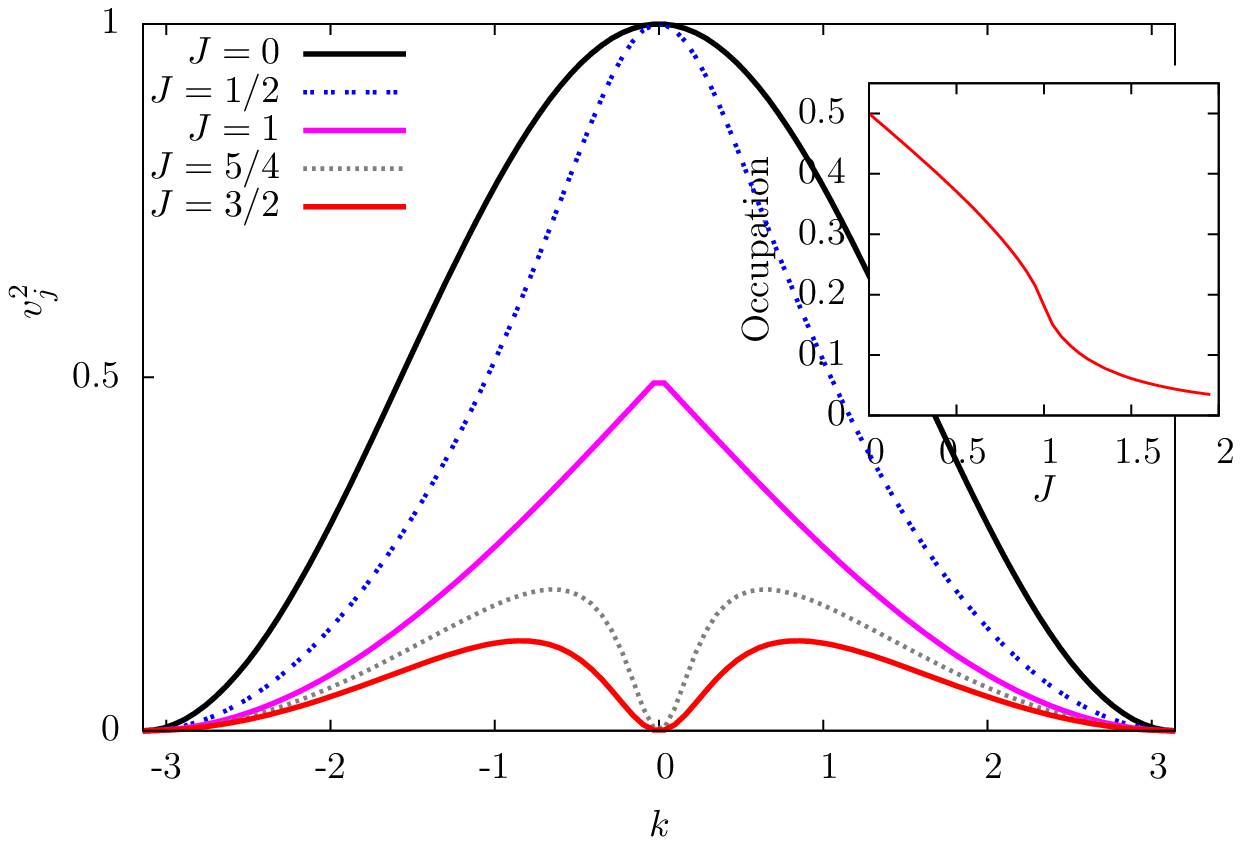,width=8cm}
\epsfig{file=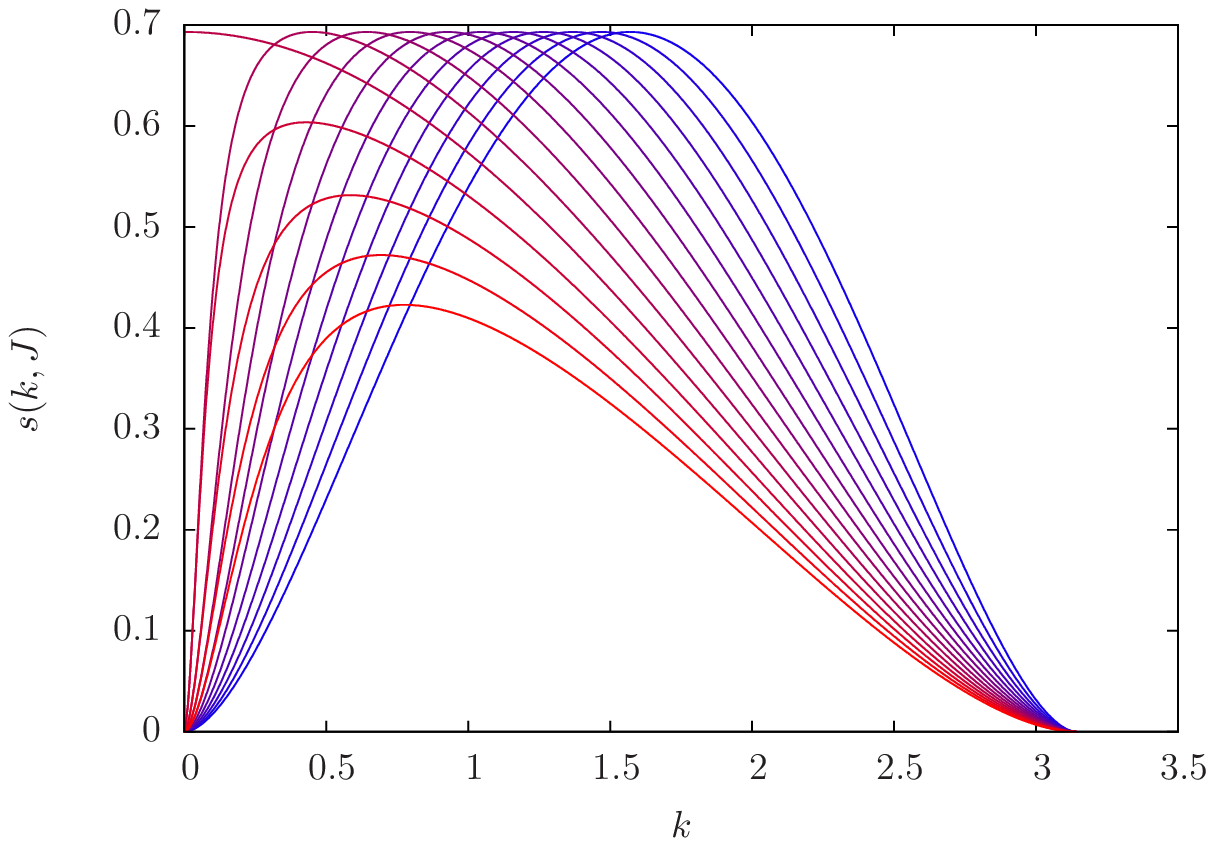,width=8cm}
\caption{\label{fig:ising}Entanglement in momentum space of the Ising
  model in a transverse field (ITF) defined in \eqref{Hxy} for
  $\gamma=1$. (A) Occupations of the different modes, $v_j^2$
  vs. $k_j$. Inset: total number of particles, as a function of
  $J$. Notice the larger slope for $J=1$.  (B) Contribution of each
  mode $k$ to the entanglement entropy, $s(k,J)=S[\varrho_j]$ for
  different values of $J$. The blue line, $J=0$, is symmetrical with
  respect to $k_c=\pi/2$. The momentum of maximal entropy,
  $k_c=\arccos J$ decreases with $J$ until, for $J=1$, it reaches
  zero. The maximal entropy contribution $s(k_c(J),J)=\log 2$ for
  $J\leq 1$. For $J>1$ (red lines), the maximal entropy is lower than
  $\log 2$.}
\end{figure}

The ground state occupations $\<d_j^\dag d_j\>$ are shown in
Fig. \ref{fig:ising} (A). As opposed to the previous case, the number
of fermions $n_{\rm f}$ (i.e., the sum of $v_j^2$ for all $j$) is not
constant, but decreases with $J$, as illustrated in
Fig. \ref{fig:ising} (A, inset). This fact indicates that the blocks
$P_n$ defined in Fig. \ref{fig:illust} have lost its special
relevance, since the Fermi point is no longer defined independently of
$J$. However, we will concentrate on the entanglement of individual
momentum pairs, $S[\varrho_j]$, and the total entanglement of positive
versus negative momenta, $P_{N/2}$.

Fig. \ref{fig:ising} (B) shows the entanglement entropy contributed by
each pair $(k_j,-k_j)$, $S[\varrho_j]$, for different values of $J$,
given by

\begin{equation}
s(k,J)=H_2\( {1\over 2} + {J-\cos k \over 2\sqrt{ (J-\cos k)^2 +
    \sin^2 k} } \)
\label{eq:entropy_itf_k}
\end{equation}

The value $J=0$ corresponds to the blue centered line. Let $k_c(J)$
denote the momentum for which $s(k_c,J)$ attains is maximum. For
$J\leq 1$, $k_c=\arccos(J)$, and $s(k_c,J)=\log 2$, i.e. that mode is
maximally entangled. For $J=1$, $k_c=0$, the maximally entangled mode
is the zero mode. For $J>1$, in the paramagnetic phase, the maximal
value $\max_k s(k,J)$ decreases with $J$ and $k_c$ increases again.

\begin{figure}[h]
\epsfig{file=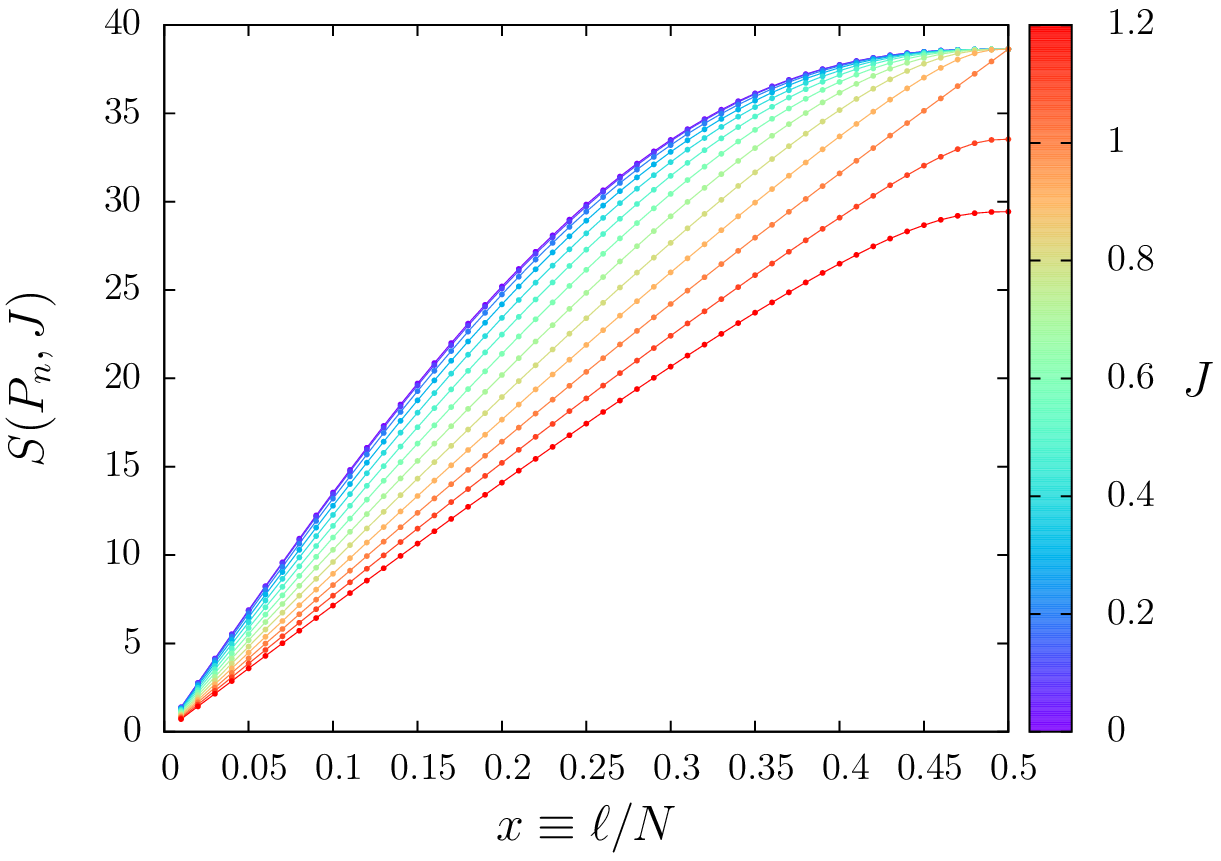,width=8cm}
\epsfig{file=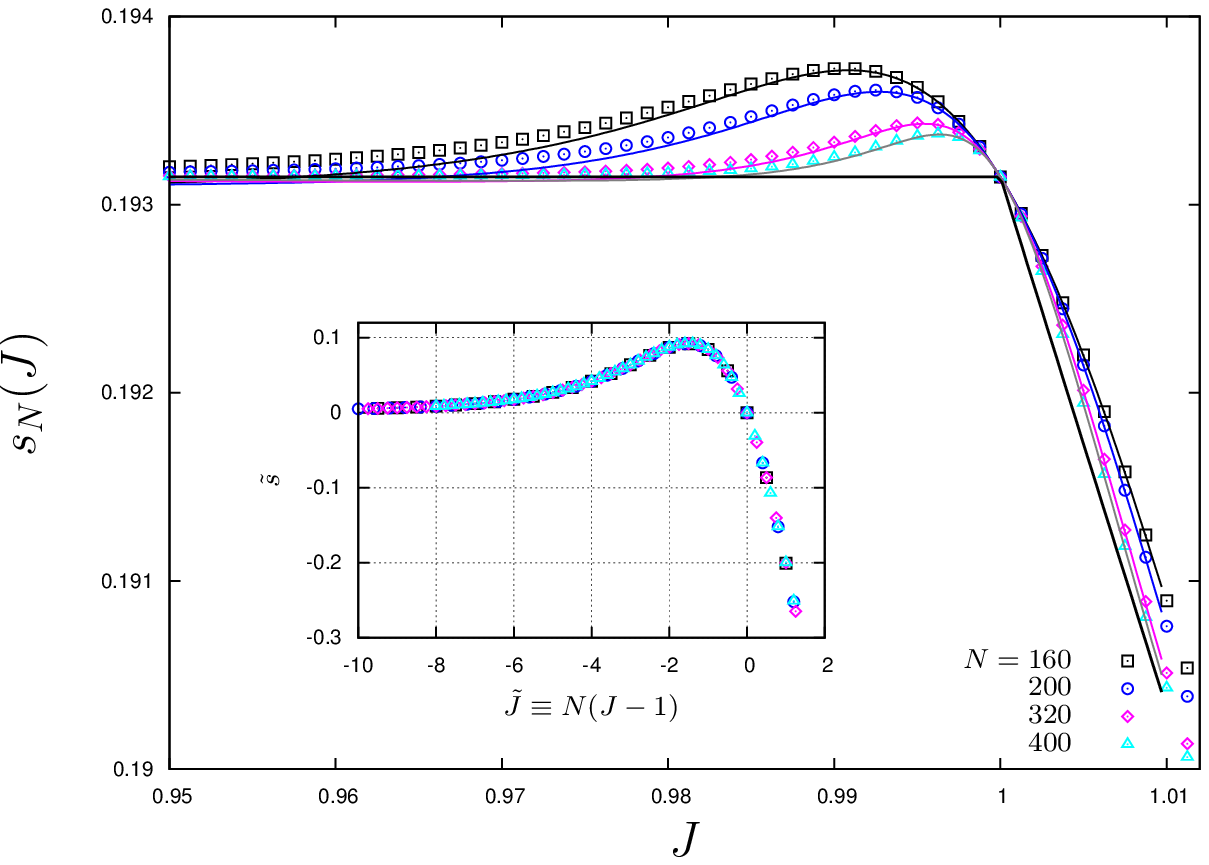,width=8cm}
\caption{\label{fig:transition}Entanglement of blocks of positive
  momenta $P_n$ ITF model, Eq. \eqref{Hxy} with $\gamma=1$. (A)
  Entropy of the $P_n$ block, $S(P_n)$, as a function of the block
  size fraction, $x=n/N$, for different values of $J$ for
  $N=200$. Notice that the maximal value of $S(P_n)$ remains constant
  for all $J\leq 1$, and decreases for $J>1$. (B) Entanglement between
  positive and negative momenta for the ITF model per site,
  $s_N(J)=S(P_{N/2},J)/N$, for different sizes. Notice how, in the
  thermodynamic limit (continuous black line), the entanglement is
  constant for $J\leq 1$ and decays linearly for $J>1$. Moreover, the
  colored continuous lines are given by approximation
  \eqref{eq:itf_single_mode}, which singles out the lowest mode.
  Inset: collapse of the finite-size curves, $\tilde s(\tilde J)$, see
  Eq. \eqref{eq:finite_size}.}
\end{figure}

In Fig. \ref{fig:transition} (A) we show the entropy of finite blocks
around the Fermi point, $S(P_n,J)$ (check Fig. \ref{fig:illust}), as a
function of $x=n/N$ for different values of $J$ in a finite system
with $N=200$. For $J=0$ the entropy is maximal for all $x$, see the
top blue curve in Fig. \ref{fig:ising} (C). The curve becomes a
straight line for the critical value, $J=1$. Notice that, for all
$J\leq 1$ the $x=1/2$ entropy is constant. In other terms, along the
ferromagnetic phase, the entanglement between the positive and
negative momenta is independent of $J$. This value, $S(P_{N/2},J)$, is
given by the expression

\begin{equation}
S(P_{N/2},J) = \sum_{0<k_j<\pi} H_2\( u_j^2(J) \) \approx {N\over\pi}
\int_0^\pi dk\; s(k,J)
\label{eq:thermodynlimit}
\end{equation}
where the last step is taken in the thermodynamic limit, and
corresponds to the area under each of the curves of
Fig. \ref{fig:ising} (B). The area under all these curves is {\em
  equal} for $J\leq 1$, and the following result can be proved:

\begin{equation}
s_0\equiv \lim_{N\to\infty} {S(P_{N/2},J)\over N} = \log 2 - 1/2 \approx
0.193 \qquad \hbox{if $J\leq 1$}
\label{eq:maxentropyitf}
\end{equation}

The point at which it is simplest to evaluate the integral is at
$J=0$, where it becomes:

\begin{equation}
s_0 = {1\over\pi} \int_0^{\pi/2} d\phi\; H_2(\cos^2(\phi))
\label{eq:integral}
\end{equation}

Fig. \ref{fig:transition} (B) shows the value of the entropy per site
between the positive and negative momenta, $s_N(J)\equiv
S(P_{N/2},J)/N$ for some finite-size values (dots) and the
thermodynamical limit (continuous black line), where we can see that
it stays constant for $J\leq 1$ and decays linearly shortly after
$J>1$. The derivative is, therefore, discontinuous at that point.

The finite-size entropies shown by the dots in
Fig. \ref{fig:transition} (B) provide very relevant information. They
all follow the scaling form

\begin{equation}
s_N(J) \approx s_0 + \tilde s(\tilde J)
\label{eq:finite_size}
\end{equation}
where $\tilde J=N(J-1)$, which can be understood as a scaling variable
since $J-1$ is the inverse of a correlation length. The inset of
Fig. \ref{fig:transition} (B) shows the collapse of the $\tilde s =
s_N(J)-s_0$ curves, when expressed as a function of $\tilde J$, in
both phases. The origin of this $\tilde s(\tilde J)$ scaling lies in
the fact that the deviation from the continuum limit is due mostly to
the the smallest momentum, $\Lambda_N=\pi/N$. If we single out its
contribution from the sum, as it is customary in the study of
Bose-Einstein condensation, we obtain the approximation

\begin{equation}
S(P_{N/2},J) \approx
H_2(u^2(\Lambda_N))+\frac{N}{2\pi}\int_{2\pi/N}^{\pi}\,d\phi\,H_2(u^2(\phi)),
\label{eq:itf_single_mode}
\end{equation}
which is shown by the colored continuous lines of
Fig. \ref{fig:transition} (B). This approximation is very accurate in
the vicinity of the phase transition, and we can see that the position
of their maxima are precisely reproduced.

\FloatBarrier

\section{Entanglement in the XXZ model}
\label{sec:xxz}

Let us now investigate the entanglement entropy in momentum space of
the XXZ model. Through the Jordan-Wigner transformation, this
corresponds to an interacting fermionic model, which we can write as

\begin{equation}
H_{XXZ} = - {1\over 2} \sum_i c^\dagger_i c_{i+1} + \mbox{h.c.} + \Delta \sum_i
n_i n_{i+1},
\label{eq:xxz_ham}
\end{equation}
where $n_i=c^\dagger_ic_i$ and endowed with anti-periodic boundary
conditions (APBC), i.e., $c^\dagger_{N+1}\equiv -c^\dagger_1$. Notice
that the number of particles is preserved in this case, so we can
restrict ourselves to the case of {\em half-filling}. This model is
known to be critical for $\Delta\in (-1,1]$, with central charge
  $c=1$. Moreover, the GS of Eq. \eqref{eq:xxz_ham} describes a {\em
    Luttinger liquid} \cite{Giamarchi}, which is characterized by
  Luttinger parameter

\begin{equation}
K= {\pi\over 2(\pi-\hbox{acos}(\Delta))}.
\label{eq:luttinger_parameter}
\end{equation}

\subsection{The N\'eel limit}
\label{sub:neel}

In the limit where $\Delta\to+\infty$ the ground state of
\eqref{eq:xxz_ham} becomes a N\'eel state, which is a superposition of
two factorized states in real space,

\begin{equation}
\ket|\Psi>={1\over\sqrt{2}}\( \ket|101010\cdots> +
\ket|010101\cdots> \)
\label{eq:neel}
\end{equation}

Let us consider the relevant momenta $\{k_j\}$, $j \in \{1,\cdots,N\}$
ordered from $-\pi$ to $\pi$ and symmetrically placed around zero, so
that $k_j=-k_{N+1-j}$. The basis states can be written as
$\ket|m_1\cdots m_{N/2};m_{N/2+1}\cdots m_N>$, with $m_j \in \{0,1\}$
the occupation of the momentum $k_j$. Then, the N\'eel state can be
written as

\begin{equation}
\ket|\Psi>=\sum_{m_1 \cdots m_{N/2}}
C_{m_1\cdots m_{N/2}}
\ket|m_1\cdots m_{N/2};\bar m_{N/2} \cdots \bar m_1>
\label{eq:neel_k}
\end{equation}
where $\bar m_j=1-m_j$, i.e. the occupation of level $k_j$ is always
the opposite of level $-k_j$. The amplitudes $C_{m_1\cdots m_{N/2}}$,
disregarding normalization, are given by:

\begin{equation}
C_{m_1\cdots m_{N/2}} = \hbox{Even}\(\sum_{i=1}^{N/2} m_i\)
(-1)^{\sum_{p=1}^{N/4} m_{2p}},
\label{eq:coeff_neel_k}
\end{equation}
where $\hbox{Even}(n)$ is defined as 1 if $n$ is even and zero
otherwise. This expression means that the non-zero wavefunction
components have an even number of particles with positive (negative)
momenta. Moreover, all the amplitudes are equal in absolute value, and
their sign is given by the parity of the occupation of the
even-indexed momenta. Notice that all modes are equally occupied,
$\<m_j\>=1/2$.

The entanglement structure of this state is as follows:

\begin{itemize}

\item{} The entropy of the positive momenta blocks $P_n$
  (Fig. \ref{fig:illust} (B)), is $S(P_n)=n \log{2}$ if $n<=N-2$.

\item{} The block which contains all positive momenta, $P_{N/2}$, is
  special, and its entropy is $S(P_{N/2})=(N-1)\log{2}$.

\item{} All the momentum pair blocks, $p_k$ which contain a pair
  $\{k,-k\}$ (Fig. \ref{fig:illust} (C)), have entropy $\log(2)$.

\item{} Blocks with two momenta which do not correspond to the same
  energy, on the other hand, present maximal entropy
  $2\log(2)$. 

\item{} Energy blocks $E_n$, Fig. \ref{fig:illust} (D), also have
  entropy $\log(2)$.

\end{itemize}

\subsection{Entanglement in Fourier space of the the XXZ model}

We have obtained numerically the GS of Hamiltonian \eqref{eq:xxz_ham}
for $N$ up to 20, and performed a numerical many-body Fourier
transform of the resulting GS, as described in section
\ref{sec:mbft}. Due to the APBC, the set of allowed momenta come
always in pairs $k_i=\pm i\pi/N$ and $i\in \{1,3,\cdots,N-1\}$. For
$L=2$ mod 4, the GS is exactly degenerate for all $\Delta$, so we
restrict ourselves to $N$ multiple of $4$.

\begin{figure}
\epsfig{file=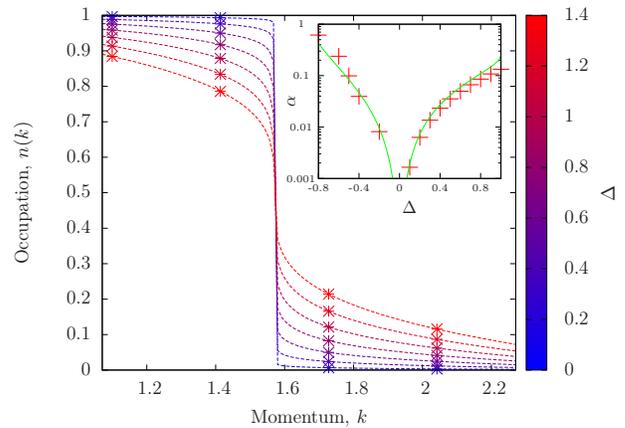,width=8cm}
\caption{Momentum space occupations for the GS of the XXZ model,
  obtained numerically for $N=20$ and several values of $\Delta$, and
  their fit to the Luttinger prediction
  \eqref{eq:occupation_luttinger}. The dependence of the occupation
  exponent with the anisotropy $\Delta$ is shown in the inset, along
  with the theoretical prediction, given by
  Eqs. \eqref{eq:occupation_luttinger} complemented with
  \eqref{eq:luttinger_parameter} and \eqref{eq:luttinger_exponent}.}
\label{fig:xxz_occup}
\end{figure}

First of all, we have investigated the occupation number of each
momentum, in order to check the known results regarding Luttinger
liquid theory. Fig. \ref{fig:xxz_occup} shows the occupation of each
$k$-mode, $n_k\equiv \bra<\Psi|b^\dagger_k b_k\ket|\Psi>$ for $N=20$
and several values of $\Delta=\{0,0.2,0.4,\cdots,1.4\}$, along with a
fit to the Luttinger liquid expression

\begin{equation}
n_k \approx |k-k_F|^\alpha.
\label{eq:occupation_luttinger}
\end{equation}
The value of the exponent $\alpha$ is related to the Luttinger
parameter $K$ through the following expression

\begin{equation}
\alpha={1\over 2} \( K + K^{-1} \) -1,
\label{eq:luttinger_exponent}
\end{equation}
and this last relation is checked in the inset of
Fig. \ref{fig:xxz_occup}.

\begin{figure*}
\epsfig{file=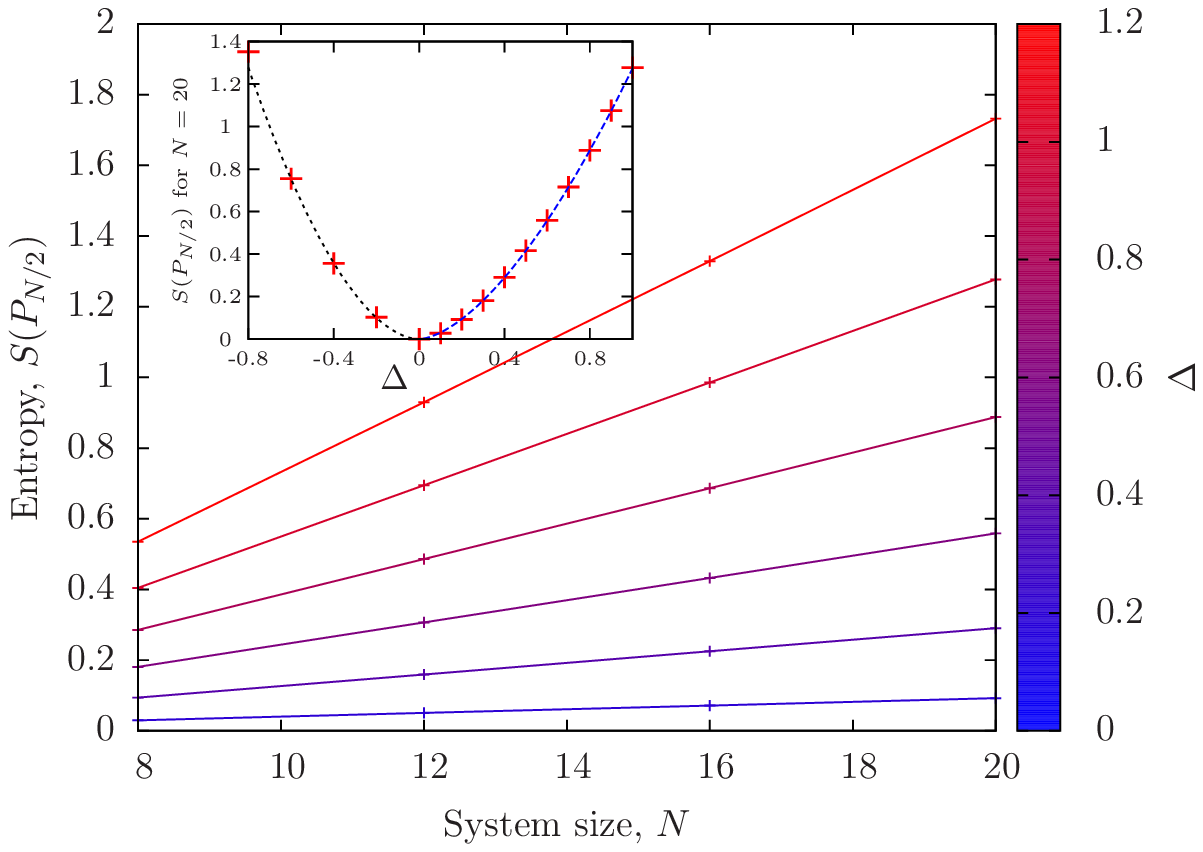,width=8cm}
\epsfig{file=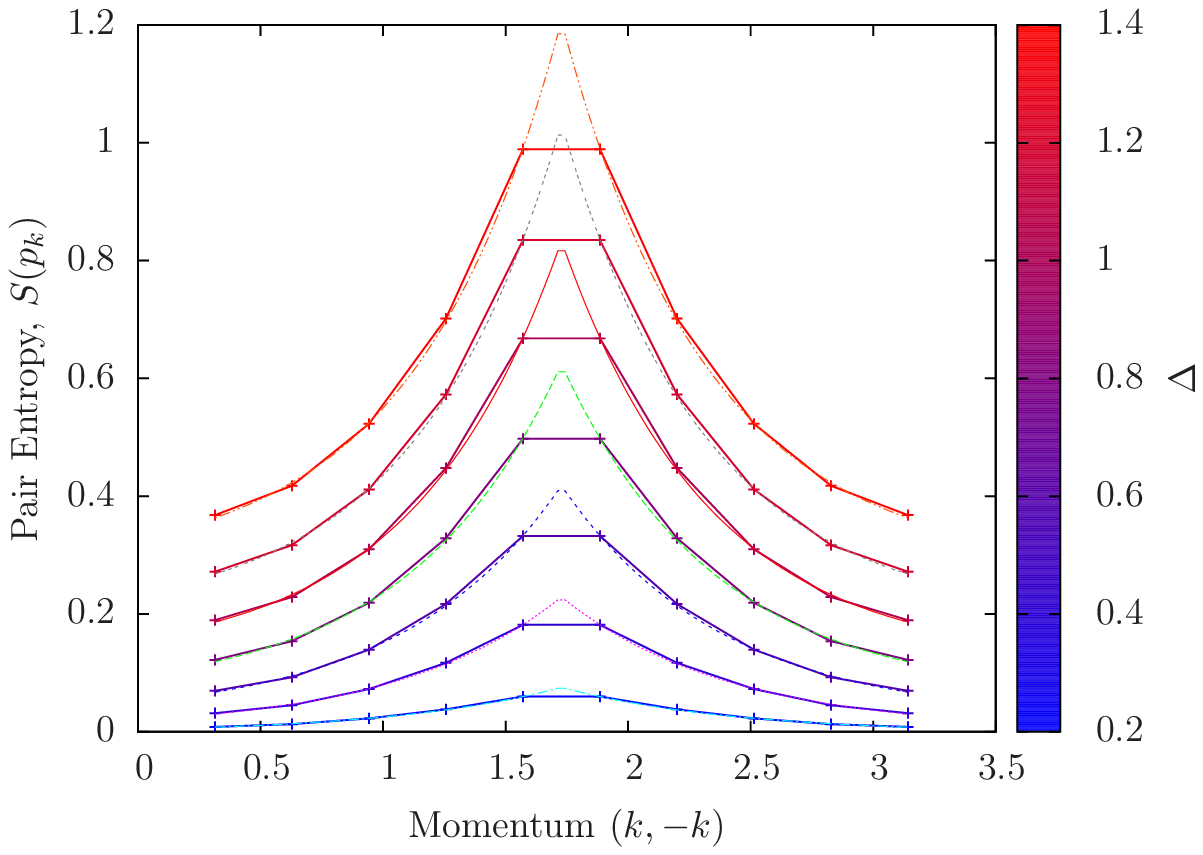,width=8cm}
\caption{(A) entropy between positive and negative momenta,
  $S(P_{N/2})$, of the GS of the XXZ model for different values of
  $\Delta$ and $N$ up to $20$. We only show positive values of
  $\Delta$ for convenience. In all cases, the plot is very accurately
  linear. Inset: $S(P_{N/2})$ with $N=20$ as a function of $\Delta$,
  along with the two different power-law fits, for $\Delta>0$, $S\sim
  \Delta^{1.62}$ and for $\Delta<0$, $S\sim \Delta^{1.83}$. (B)
  Entropy of the block containing only an opposite pair of momenta,
  $S(p_k)$ as a function of $k$, which takes the maximal value for
  $k\approx k_F$. In all cases, the entropy decays as a power-law of
  the distance to the Fermi momentum.}
\label{fig:xxz_sk}
\end{figure*}

The entropy of the positive vs negative momenta, $S(P_{N/2})$ (see
Fig. \ref{fig:illust}) is shown in Fig. \ref{fig:xxz_sk} (A) as a
function of the system size, $N$ for several values of $\Delta$, where
we show only positive values of $\Delta$ for clarity. The dependence
with the system size is linear for all values of $\Delta$. The inset
shows the dependence of the entropy between positive and negative
modes with $\Delta$ for $N=20$. A very good fit can be made to a power
law, with a different exponent for positive and negative $\Delta$:
$S(P_{N/2}) \sim \Delta^{1.62}$ for $\Delta>0$ and $\Delta^{1.83}$ for
$\Delta<0$.

The right panel, Fig. \ref{fig:xxz_sk} (B), shows the entropy of
blocks of the form $p_k$ (see Fig. \ref{fig:illust}), which contain
only a pair of opposite momenta, as a function of $k$, for $N=20$ and
several values of $\Delta>0$. Notice that, in the XY and ITF model,
all those entropies were zero. In all cases, this entropy is highest
for $k\approx k_F$, and decays exponentially away from the Fermi
point. The fits in Fig. \ref{fig:xxz_sk} (B) are done to an expression
of the form

\beq 
S_k \approx S_{k_F} \exp(-|k-k_F|/\sigma) + S_0,
\label{eq:fit_pair_entropy}
\eeq
where $\sigma$ provides a measure of the extent to which momenta away
from the Fermi surface can be removed without alteration of the rest
of the state. This $\sigma$ parameter grows as $|\Delta|$ does.

\begin{figure}
\epsfig{file=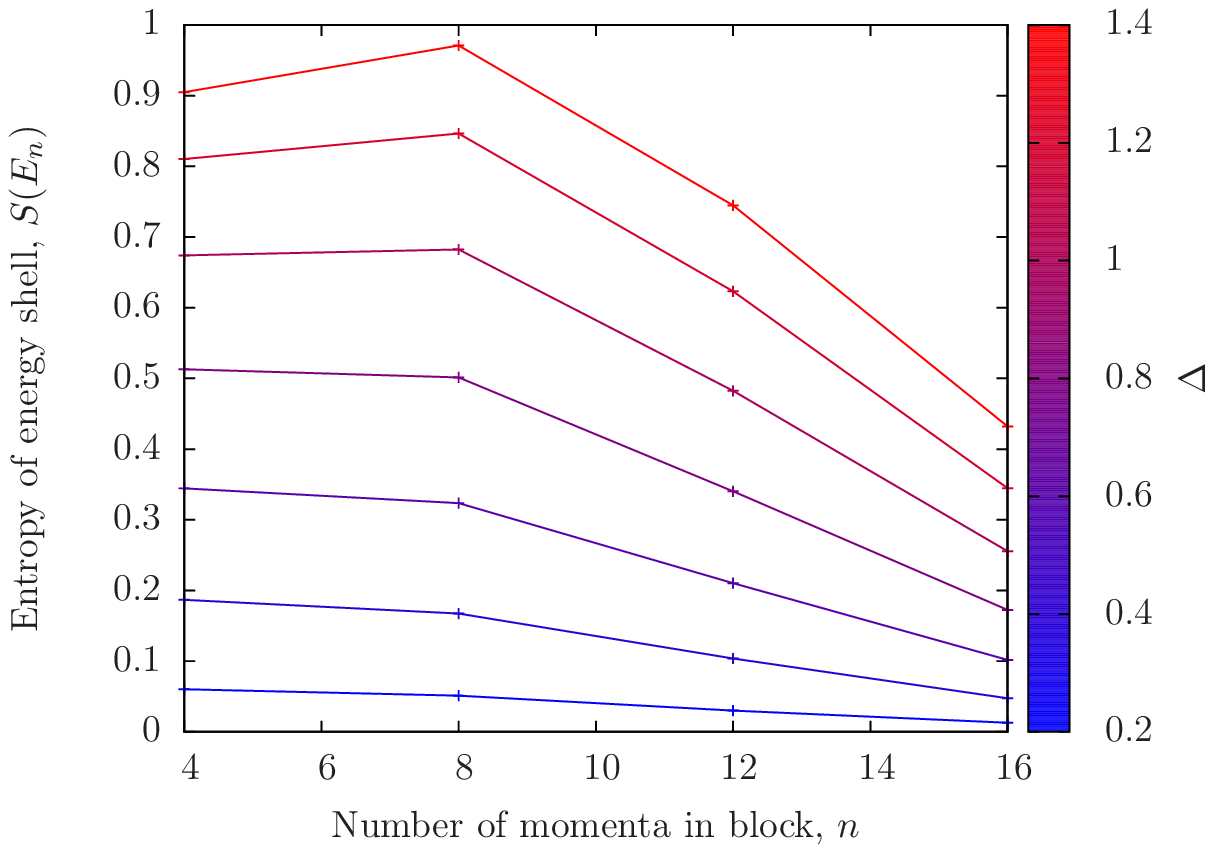,width=8cm}
\epsfig{file=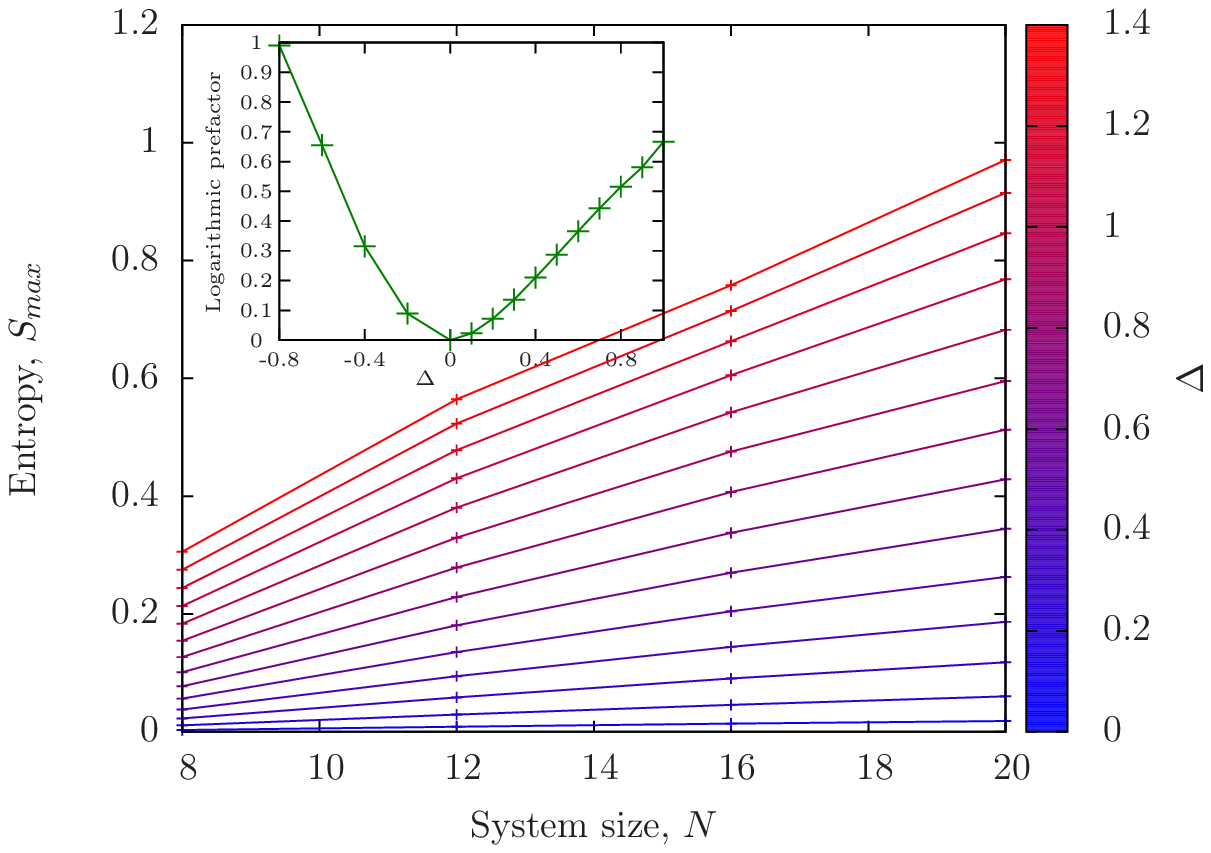,width=8cm}
\caption{(A) Entanglement entropy of blocks formed by the $n$ energy
  levels closest to the Fermi energy, $E_n$ (see
  Fig. \ref{fig:illust}), for different positive values of $\Delta$,
  also for $N=20$. (B) Maximal entanglement entropy of the $E_n$
  blocks, $S_{\rm max}$ as a function of $N$, for different values of
  positive $\Delta$. The best fit is to a logarithmic growth,
  \eqref{eq:xxz_finitesize} with $N^{-2}$ corrections. Inset:
  prefactor of the logarithmic term as a function of $\Delta$, both
  positive and negative. In all cases, the fit error is less than
  $10^{-7}$ for $\Delta \in (-1,1)$.}
\label{fig:xxz_energy}
\end{figure}

We have also studied the blocks containing the $n$ momenta which are
closest to the Fermi energy, denoted by $E_n$ in
Fig. \ref{fig:illust}. Fig. \ref{fig:xxz_energy} (A) shows the entropy
$S(E_n)$ as a function of $n$ for $N=20$ for different positive values
of $\Delta$. This case bears the strongest similarity to the energy
blocks studied previously in \cite{Laguna.PRB.14}. It can be noticed
that the value $S(E_n)$ always decreases for large $n$, showing that
separating the momenta which are further away from the Fermi surface
always has a smaller entropic cost. The maximal value of $S(E_n)$,
$S_{\rm max}$ is always found for small values of $n$. 

The panel (B) of Fig. \ref{fig:xxz_energy} shows the maximal entropy
$S_{\rm max}$ as a function of $N$ for different positive values of
$\Delta$. When $\Delta$ is in the critical region, $\Delta\in (-1,1]$,
  the best fit is always to the form

\begin{equation}
S_{\rm log}(N)=\Theta \log(N) + \beta + \gamma/N^2,
\label{eq:xxz_finitesize}
\end{equation}
i.e., a logarithmic growth with a finite-size correction of
$N^{-2}$. Outside the critical region, and even for $\Delta=1$, the
best fit is not logarithmic, but to a power-law growth. The inset of
Fig. \ref{fig:xxz_energy} (B) shows the logarithmic growth prefactor,
$\Theta$, as a function of $\Delta$. It presents a minimum,
$\Theta=0$, for the XX case, $\Delta=0$, and appears to grow linearly
for positive $\Delta$.


\section{Conclusions}

The structure of a many-body quantum state can be studied by {\em
  slicing} the Hilbert space in different ways. Traditionally, we
consider the entanglement of real space blocks, but in some cases we
can gather useful information using other bases, such as Fourier
space. The success of DMRG in momentum space for different models
points at a slow growth of entropy with the system size if the path is
chosen appropriately. In this work we have studied the entanglement
structure in Fourier space of spin chains, described as fermionic
states through the Jordan-Wigner transformation. We have benefitted
from both analytical and numerical tools in our study.

First, we have studied the generalized XY model, which can be
analytically solved using a Bogoliubov transformation. In all cases,
momentum $k$ is only coupled to its conjugate $-k$. For the XX model,
the wavefunction is factorizable. Introducing the $\gamma$ parameter,
which measures the anisotropy between the $X$ and $Y$ axes, we couple
the momenta pairs, which for $\gamma\to\infty$ become maximally
entangled. The Ising model in a transverse field (ITF) can also be
studied analytically. In that case, we find that the entanglement
entropy per site between the positive and negative momenta is constant
in the ferromagnetic phase and decreases linearly with the external
field in the paramagnetic phase for $J$ close to $J=1$. Thus, there is
a clear signature of the quantum phase transition. Near the critical
point, the finite size corrections to the entropy can be obtained by
singling out the contribution of the lowest momentum.

We have also studied the XXZ model numerically, with sizes up to
$N=20$. In this case, the most salient feature is that the maximal
entropy among the energy-space blocks $E_n$, which contain the $n$
closest momenta to the Fermi energy, grows with $\log(N)$ in the
critical region. The prefactor, nonetheless, is not related to the
central charge, but instead it depends on the compactification radius,
which is given by the anisotropy parameter $\Delta$. Out of the
critical region, the entropy of energy blocks grows faster than
logarithmically, very likely as a power-law. The N\'eel state, which
is the limit for infinite anisotropy, has also very peculiar features
in Fourier space, such as a constant entropy for all of energy blocks.

Is there a complementarity relation between entanglement in real and
momentum space? The question is even difficult to formulate
rigorously, due to the inherent inhomogeneity of momentum space. For
example, which blocks should be used for the comparison? One may
characterize the maximal entanglement $S_M$ in momentum space using a
{\em minimax} definition. Let ${\cal P}$ be the set of all {\em
  permutations} of the (momentum) sites. Given
$p=\{s_1,\cdots,s_N\}\in {\cal P}$, we may obtain the maximal entropy
among the blocks starting from $s_1$, and define:

\begin{equation}
S_M \equiv \min_{\cal P}\; \max_m\; S(\{s_1,\cdots,s_m\}).
\label{eq:minmax}
\end{equation}
(See \cite{Laguna.07} for a similar idea in the application of the
DMRG on networks). Equivalently, we can consider the minimal entropy
among all blocks of a given size $m$, and maximize on $m$:

\begin{equation}
S_M \equiv \max_n\; \min_{|B|=n}\; S(B).
\label{eq:minmax}
\end{equation}

Both expressions must yield the same result. For example, for a 1D
translation-invariant system in real space, the minimax entropy block
will contain no holes, and the optimal permutation will follow the 1D
structure. For the XXZ model in momentum space we conjecture that the
minimax entropy takes place for the energy blocks around the Fermi
surface. Thus, we propose to investigate the relation between these
minimax entropies in real and momentum space.

Moreover, it is relevant to ask whether the signatures for critical
behavior in Fourier space that we have discussed can be summed up into
a universal criterion, at least for 1D systems. Numerical
investigation of the many-body Fourier transform is very demanding
computationally, thus a conceptual breakthrough is necessary at this
step.


\begin{acknowledgments}

We would like to thank J. Dukelsky and F. Alcaraz for useful
discussions. This work was funded by grants FIS-2012-33642 and
FIS-2012-38866-C05-1, from the Spanish government, QUITEMAD+
S2013/ICE-2801 from the Madrid regional government and SEV-2012-0249
of the ``Centro de Excelencia Severo Ochoa'' Programme.

\end{acknowledgments}


\end{document}